\documentclass[fleqn,10pt]{wlscirep}
\usepackage[utf8]{inputenc}
\usepackage[T1]{fontenc}
\usepackage{xcolor}

\title{Echo chambers and information transmission biases in homophilic and heterophilic networks}

\author[1]{Fernando Diaz-Diaz}
\author[1]{Maxi San Miguel}
\author[1,*]{Sandro Meloni}

\affil[1]{IFISC (UIB-CSIC), Institute for Cross-Disciplinary Physics and Complex Systems, Campus Universitat de les Illes Balears E-07122, Palma de Mallorca, Spain}

\affil[*]{sandro@ifisc.uib-csic.es}

\begin{abstract}

We study how information transmission biases arise by the interplay between the structural properties of the network and the dynamics of the information in synthetic scale-free homophilic/heterophilic networks. We provide simple mathematical tools to quantify these biases. Both Simple and Complex Contagion models are insufficient to predict significant biases. In contrast, a Hybrid Contagion model -in which both Simple and Complex Contagion occur- gives rise to three different homophily-dependent biases: emissivity and receptivity biases,and echo chambers. Simulations in an empirical network with high homophily confirm the existence of these biases. 
Our results shed light into the mechanisms that cause inequalities in the visibility of information sources, reduced access to information, and lack of communication among distinct groups.

\end{abstract}
\begin{document}

\flushbottom
\maketitle

\thispagestyle{empty}
\noindent 

\section*{Introduction}
Information transmission in the context of Information and Communication Technologies is a great opportunity to create a better-informed society, but in practice these technologies are also promoting phenomena such as viral spreading of fake news \cite{goel_structural_2015, vosoughi_spread_2018, juul_comparing_2021}, echo chambers \cite{cinelli_echo_2021, baumann_modeling_2020, baumann_emergence_2021}, perception biases such as false consensus or majority illusions \cite{lee_homophily_2019}, as well as social polarization \cite{ baumann_modeling_2020, baumann_emergence_2021,tokita_polarized_2021}. We understand by echo chambers situations in which the transmission of information among individuals belonging to the same opinion group is dominant, while transmission among individuals with different opinions is hindered. The real social impact of echo chambers and its causal link with misinformation cascades are debated topics \cite{dubois_echo_2018, barbera_tweeting_2015,tornberg_echo_2018}, but data-driven and computational approaches confirm that the structural properties of social networks are tied to the emergence of echo chambers \cite{cinelli_echo_2021, baumann_modeling_2020, baumann_emergence_2021}. In particular, the homophily of the network -that is, the tendency of nodes to be connected to other nodes of the same group- seems to be a key ingredient to generate echo chambers \cite{baumann_modeling_2020} and perception biases \cite{lee_homophily_2019}.


Among the phenomena associated with information transmission, echo chambers are presently the subject of an intensive research, but they are only an aspect of a broader subject:  
the "information transmission biases" (in short, IT biases), which include all the possible alterations in the transmission of information that appear when changing the nature of the nodes that generate and receive such information. In addition to echo chambers, examples of IT biases include: a) An enhanced/inhibited emission of information by a certain group (for example,  how female opinions were wronged and overheard based on gender stereotypes \cite{fricker_epistemic_2007}); and b) an enhanced/inhibited reception of information by a certain group. These biases in information transmission have been observed in real-word networks and are influenced by their structural properties \cite{mesoudi_bias_2006}. Let us remark that some studies regarding IT biases focus on how distinct types of information have different transmission probability \cite{altshteyn_evidence_2014, fay_socially_2021}, while here we will focus on the differences in the transmission induced by intrinsic properties of the node that emits and/or receives the information in the network. \\

Modeling processes of information transmission first requires the choice of a dynamical model. Often, the spreading of information is assumed to follow the same laws as the spreading of a disease. Because of this, epidemic models (also called \textit{Simple Contagion} models) \cite{pastor-satorras_epidemic_2015} have been used for discussing transmission of information. However, spreading of information, adoption of innovations etc. are examples of social contagion phenomena in which individuals often require multiple exposure to a given information to adopt it \cite{chang2018}. These social mechanisms are included in models of \textit{Complex Contagion} \cite{centola_cascade_2007,centola2018,gleeson_cascades_2008,hackett_cascades_2011,oh_complex_2018} inspired in the Threshold Model by Granovetter \cite{granovetter_threshold_1978,watts_simple_2002} in which adoption requires a threshold number of neighboring agents that have already adopted it. In this sense, Complex Contagion, at variance with Simple Contagion, is a nonlinear process that  requires group or many-agent interactions. Still, these group interactions are built from a combination of pairwise interactions, while possible higher order many-agent interactions would call for a different approach \cite{iacopini_simplicial_2019,battiston2020,lambiotte2021,de_arruda_multistability_2021}. Several works have addressed the question of the validation of Complex Contagion models against experimental data \cite{centola_spread_2010,couzin2015,aral2017}, as well as the comparison of Simple and Complex Contagion models in this empirical context \cite{lerman2016,monsted2017,centola2021}. However, there is also empirical evidence that many processes of information transmission involve both Simple and Complex Contagion, with some agents adopting in a single interaction and others requiring multiple exposures \cite{state2015,kertesz2016}. As a consequence, different models of \textit{Hybrid Contagion} combining Simple and Complex Contagion have been proposed \cite{dodds_universal_2004,czaplicka_competition_2016,kertesz2016,min_competing_2018,kook_double_2021}. In this paper we will compare predictions of Simple, Complex and Hybrid Contagion models concerning IT biases.

A second ingredient in the modeling of information transmission is the choice of an underlying social network. In this work we do not construct a network which, for example by a community structure,  is expected a priori to lead to IT biases. Instead, we focus on how echo chambers and other IT bias can emerge from the interplay between the structural properties of the network and the dynamics of the information.
An important issue in this problem is disentangling the effects of social influence, included in the dynamical models, and homophily. To this end we choose here to consider Barabasi-Albert \cite{barabasi_emergence_1999} networks with tunable homophily, following the model proposed in \cite{karimi_homophily_2018} for social networks. 


In this paper we provide mathematical tools that can be used to define IT biases and to find out in which parameter regimes those IT bias exist. We show that the mechanism of \textit{Hybrid Contagion} leads to the three IT biases mentioned before:  emissivity bias, receptivity bias and echo chambers. Importantly, the echo chamber bias, which is not present in neither Simple nor Complex Contagion, arises for a wide range of  homophily parameters for Hybrid Contagion. Moreover, simulations in an empirical network confirm the presence of echo chambers and other IT biases. 


%
%
%
%
%
%
%

\section*{Results}

\subsection*{Homophilic and heterophilic networks}
\begin{figure}[ht]
\centering
\includegraphics[width=0.8\linewidth]{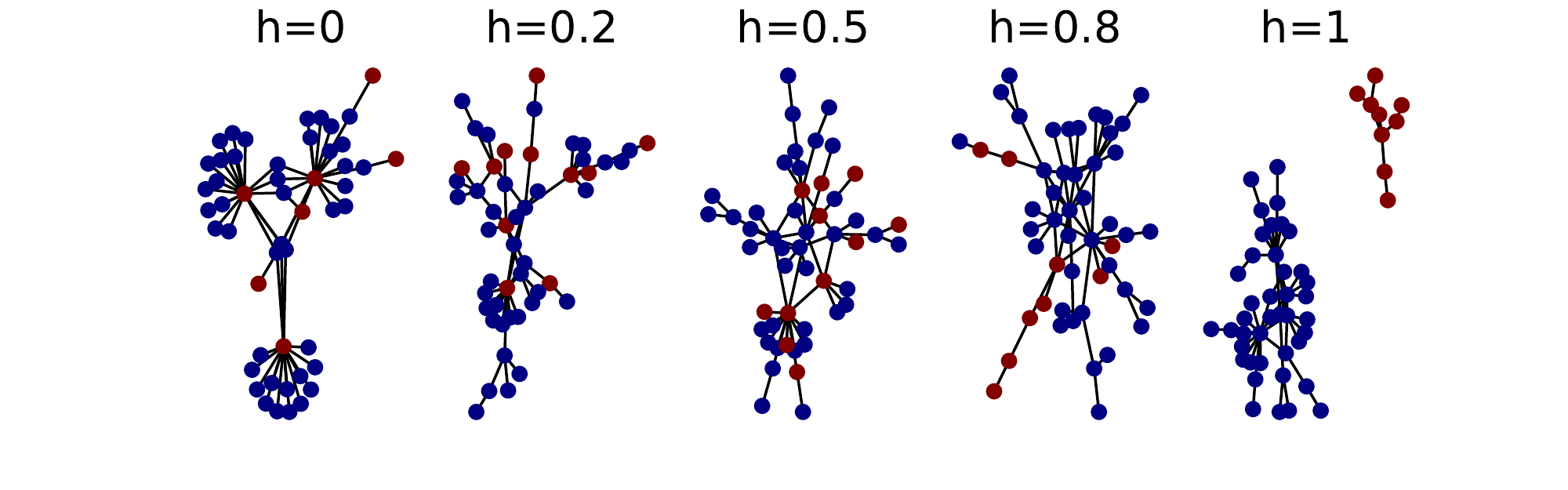}
\caption{Topology of the networks generated by the BAh model, for different values of the homophily parameter $h$. Blue nodes indicate majority group, while red nodes indicate minority group. Other parameters: number of nodes $N=50$, minority fraction $f_a=0.2$. }
\label{aux1}
\end{figure}

To accurately model the structure of social networks, we use the Barabasi-Albert-homophily model (BAh model) proposed by Karimi et al. \cite{karimi_homophily_2018, lee_homophily_2019 }. This model is a generalization of the classical Barabasi-Albert (BA) model \cite{barabasi_emergence_1999} for scale-free networks.
In this model, each node has a binary attribute, called the \textit{group} of the node. We distinguish a \textit{majority} and a \textit{minority} group. The fraction of minority nodes in the network, i.e., the probability that a newly introduced node belongs to the minority group, is denoted by $f_a(<0.5)$. As in the classical BA model, one node $i$ is added to the network and connected to $m$ existing nodes in each time step. Unlike it, however, the probability of attachment with an existing node $j$, $\Pi_{ij}$, depends not only on the degree of the existing node, $k_j$, but also on a \textit{homophily parameter} $h$:
\begin{align}
 \Pi_{ij}&\propto
\begin{cases}
    h k_{j} \ \ \ \ \ \ \ \ \ \ \ \ &\textrm{if} \ \ \ \ \textrm{Group($i$)$=$Group($j$)} \\
   (1-h) k_{j} \ \ &\textrm{if} \ \ \ \textrm{ Group($i$)$\neq$Group($j$)}
    \end{cases} \label{prob_hom}
\end{align}
From equation \eqref{prob_hom}, it is clear that setting $h>0.5 \ (h<0.5)$ generates homophilic (heterophilic) networks. For $h=0.5$, one recovers the standard Barabasi-Albert model; while for $h=1$, the probability of connecting two nodes of different groups is zero; thus, the network fragments into two components corresponding to the two groups. Figure \ref{aux1} shows examples of networks generated using this model, for different values of the homophily parameter $h$. A first insight we can take is that in the heterophilic regime (i.e., $h<0.5$), the hubs are minority nodes, while in the homophilic regime (i.e., $h>0.5$) they belong to the majority group. Additionally, for $h<0.5$ the average degree of the hubs is higher than for $h>0.5$. This happens because in heterophilic networks, the abundant majority nodes preferably attach to the rare minority ones, leading to larger degrees. In contrast, in homophilic networks the probability of choosing one particular majority node to make a connection is smaller, due to their higher abundance. As will be shown below, this disparity in the hubs' sizes influences the information transmission on the different networks.

\subsection*{Simple vs Complex Contagion}

\begin{figure}[ht]
\centering
\includegraphics[width=0.8\linewidth]{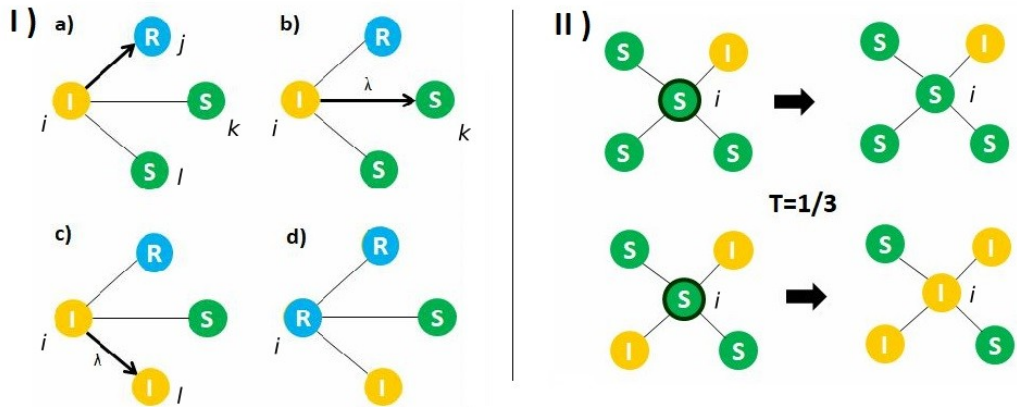}
\caption{ (I) Simple Contagion: each informed node (I) tries to inform all of its susceptible (S) neighbours before becoming recovered (R). a) Node $i$ tries to inform node $j$, but fails because $j$ is in a recovered state. b) Then, node $i$ tries to inform node $k$, with probability of success $\lambda$, failing in this case. c) Finally, node $i$ succeeds in informing node $l$. d) After having tried to inform all of its neighbours, node $i$ becomes recovered (R). \\
(II) Complex Contagion (with threshold $T = \frac{1}{3}$). Upper panels: node $i$ has only a fraction $\frac{1}{4}$ of informed neighbours, so it remains susceptible. Lower panels: Node $i$ has a fraction $\frac{1}{2}$ of informed nodes, so the threshold is surpassed and therefore it becomes informed.}
\label{aux2}
\end{figure}

\subsubsection*{Definitions and description of the models}
We want to study the differences in information transmission that arise when considering different emitters and receivers of information. We call \textit{source node} the node that initiates the information transmission (i.e., the seed of the transmission process). We also define the group to which it belongs as the  \textit{source group}. If the source node belongs to the majority, we talk about \textit{majority source}, otherwise we call it \textit{minority source}. \\
To quantify the spreading of information between groups, we define four information transmission observables (in short, \textit{IT observables}) denoted by $IT_{ab}$. Each one represents the probability of successfully transmitting information from a source of a given group $a$ to a target of a given group $b$, where $a$ and $b$ can be minority ($m$) or majority ($M$). For example, the probability of information transmission from a minority source to a minority target is represented by $IT_{mm}$, the probability from minority source to majority target is represented by $IT_{mM}$, and so on. In addition, please note that calculating one $IT_{ab}$ is equivalent on average to finding the final density of informed nodes belonging to the group $b$, when the seed of the contagion belongs to group $a$. The details of this equivalence and other aspects of the simulation procedure can be found in the Methods section. \\

To start modelling information transmission (IT) on homophilic and heterophilic networks, we first consider a Simple Contagion model used by Karimi et al. \cite{karimi_homophily_2018}.This is a  modification of the SIR dynamics \cite{pastor-satorras_epidemic_2015} in which each node can be in one of the following three possible states: susceptible, informed (also called adopter) and recovered. In this context, "recovered" refers to nodes that know the information, but choose not to spread it. When a node becomes informed, it tries to propagate the information to each of its neighbours only once. In each of these trials, recovered nodes cannot become informed again, while susceptible nodes become informed with probability $\lambda$. This parameter $\lambda$ is called the \textit{infectivity} of the contagion process. Once an informed node has tried to inform all of its neighbours, it automatically becomes a recovered node. The main difference with the standard SIR model is that, in the considered model, each link has only one opportunity to transmit the information, while in the standard SIR model each node will continue to transmit the information until it transitions into the recovered state. Sketches of the transitions can be found in the left panels of figure \ref{aux2}. \\ 

We will compare the results of the Simple Contagion model with processes of Complex Contagion \cite{centola_cascade_2007,centola2018}. For Complex Contagion we use the threshold model proposed by Granovetter \cite{granovetter_threshold_1978} and later studied by Watts \cite{watts_simple_2002}. In this model, in each time step one node of the network is randomly selected. If the node is informed, it remains in that state. If the node is susceptible, it changes state if the fraction of informed nodes in its neighborhood exceeds a value $T$, called the \textit{threshold} of the Complex Contagion. An illustration of this process can be found in the right panel of figure \ref{aux2}.\\ Complex Contagion gives rise to cascade processes that manifest themselves in first order transitions, i.e., a discontinuous jump in the global maximum of the probability density function (pdf) of the density of informed nodes. In contrast, a Simple Contagion model exhibits a second order transition, i.e, the maximum of the pdf changes continuously. Notice that the control parameters of both models have different meanings and opposite behavior: in Complex Contagion, a larger threshold $T$ inhibits contagion; while in Simple Contagion, a larger infectivity $\lambda$ enhances it.

\subsubsection*{Capturing the impact of homophily in information spreading by using Complex Contagion}
\begin{figure}[ht]
\centering
\includegraphics[scale=0.75]{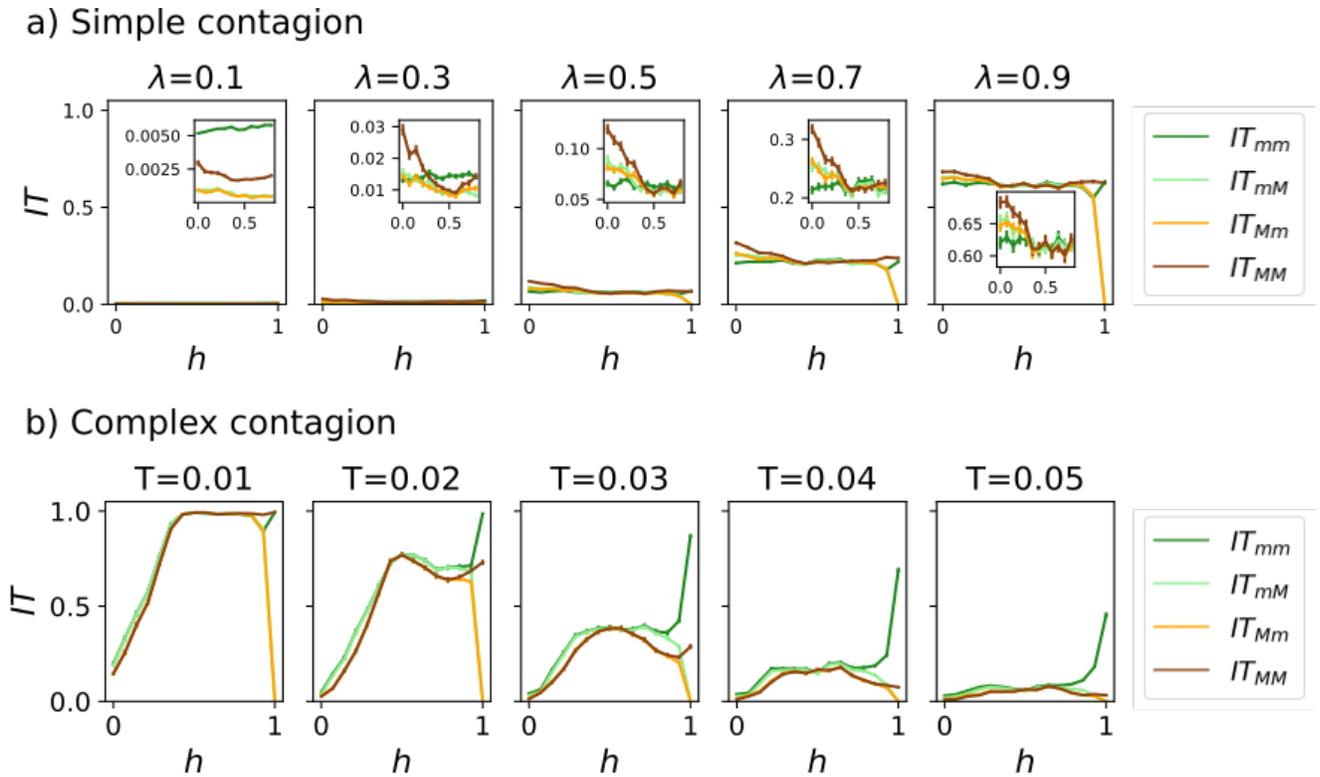}
\caption{Information transmission observables as a function of the homophily parameter $h$. The chosen dynamics are Simple Contagion in panel a and  Complex Contagion in panel b. Each column shows a different infectivity $\lambda$ or threshold $T$. The inset in panel a depicts the same IT observables as a function of the homophily parameter, but with a rescaled y axis, so that variations among the curves are visible. To improve clarity, we have removed the values corresponding to $h>0.85$ from the inset. In the legend, the subindexes indicate source and target groups respectively (for example, $IT_{mM}$ indicates that the source group is the minority and the target group is the majority). Parameters used: $N=1000$ nodes, minority fraction $f_a=0.2$, $M=1000$ realizations (varying both the network structure and the location of the seed).  }
\label{fig1}
\end{figure}

Let us use the defined IT observables to compare Simple and Complex Contagion. In figure \ref{fig1}, we show how the change of the homophily parameter $h$ affects information transmission. For Simple Contagion, the dependency of our four observables on the homophily parameter is very weak (panel a). The strongest dependence appears for $h=1$, but it is trivial because the network is fragmented and there is no path connecting different groups.  When zooming in (insets of panel a), slight dependencies in both the homophily parameter and the source and target groups can be noticed. These differences were discussed by Karimi et al. \cite{karimi_homophily_2018}. Changing the degrees of the source and target nodes does not cause novel behaviors, so they can be disregarded as control parameters (not shown). Importantly, this simplified model does not account for the observed strong biases in the presence of homophily \cite{lee_homophily_2019, cinelli_echo_2021}. \\
On the other hand, for Complex Contagion, we observe a stronger dependency on the homophily parameter $h$ (figure \ref{fig1}, panel b). In fact, we find a completely different behavior in the homophilic and heterophilic regimes. As long as $h<0.5$, increasing $h$ improves information transmission. This is caused by the larger degree of the heterophilic hubs (fig. \ref{aux1}), which prevents Complex Contagion \cite{watts_simple_2002, centola_cascade_2007}. On the other hand, for $h>0.5$, the effect of homophily is much less pronounced, because the change in the hubs' degree is much milder in the homophilic regime. Additionally, increasing the threshold decreases all IT observables, although the decrease is more pronounced when $h>0.5$. There are also slight dependencies on the source group but not on the target group (except for $h\approx 1$), suggesting some kind of bias in the transmission process.

\subsubsection*{Critical threshold for Complex Contagion}
After analyzing the role of the homophily parameter, a question naturally arises: does $h$ affect the nature of the transition? In other words, will we find a first order transition with an "all-or-nothing" outcome, so that either every node becomes informed or only a negligible fraction does? \\
\begin{figure}[ht]
\centering
\includegraphics[scale=0.7]{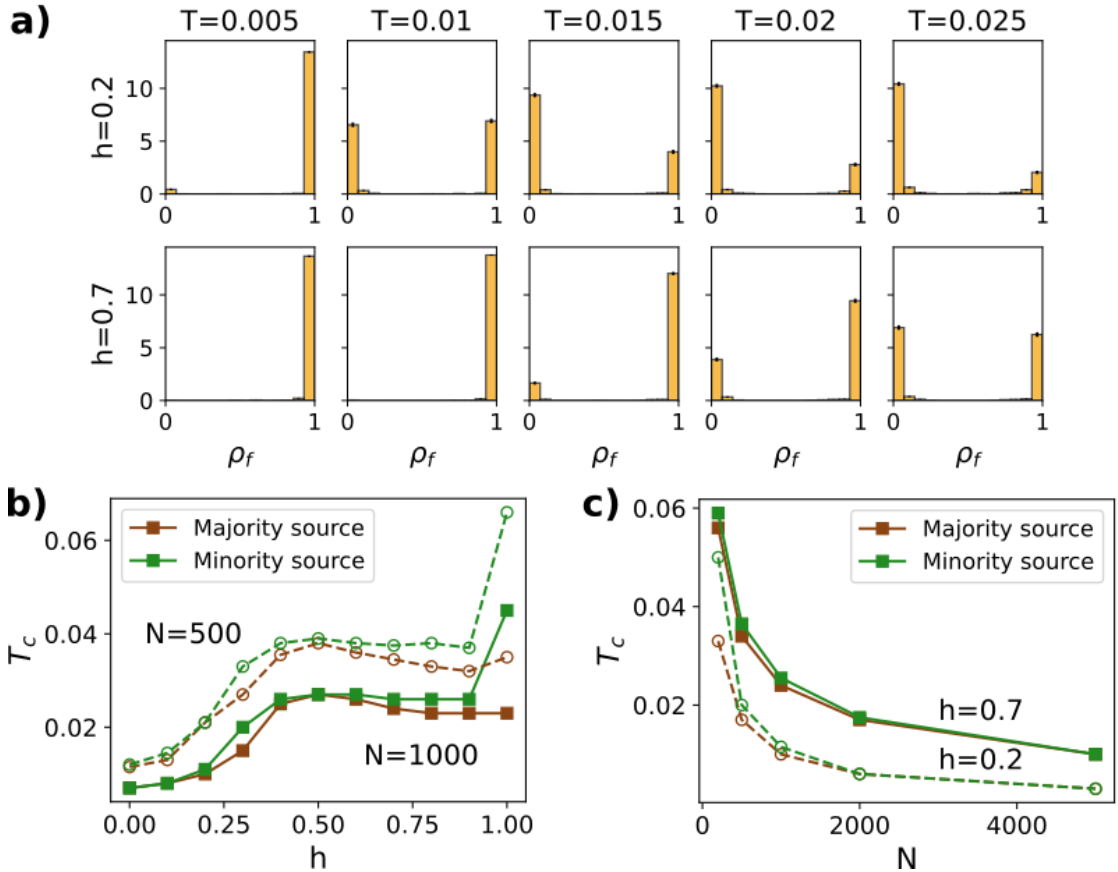}
\caption{ Critical threshold for Complex Contagion dynamics. a): Histograms with the probability distribution of the final density of informed nodes, for a majority source and $h=0.2$ (upper row) and $h=0.7$ (lower row). The pdf is normalized so that the total area of each histogram equals one.  Used parameters: $N=1000$ nodes, $M=1000$ realizations. b): Critical threshold as a function of the homophily parameter, for minority and majority sources. Simulations were performed for two values of $N$: $N=500$ (dashed line) and $N=1000$ (solid line). c): Critical threshold as a function of the number of nodes, for minority and majority sources and two values of the homophily parameter: $h=0.2$ (dashed line) and $h=0.7$ (solid line). In all simulations, the minority size was fixed to $f_a=0.2$.
     }
\label{fig2}
\end{figure}
To answer this question, we study how the probability density function (pdf) of the density of informed nodes changes when varying the threshold $T$, for two values of $h$ (figure \ref{fig2}). The histograms confirm the existence of a first order transition in both cases: the pdf shows two maxima at $\rho_f=0$ and $\rho_f=1$, corresponding to no contagion and full contagion respectively. \\
The histograms allow to identify the critical threshold $T_c$, which is defined as the point where the global maximum changes from $\rho_f=0$ to $\rho_f=1$. Calculating the critical threshold for varying levels of homophily, we obtain the plot from panel \ref{fig2}b, where nontrivial dependencies on $h$ arise. In the homophilic regime ($h>0.5$), $T_c$ is almost constant, while in the heterophilic regime ($h<0.5$), $T_c$ decays as the network becomes more heterophilic ($h\to0$). This means that information propagates less efficiently in heterophilic networks, in agreement with figure \ref{fig1}b.  Moreover, for $h=1$, the threshold for minority sources increases significantly. This is a consequence of the fragmentation of the network: the minority group forms a separate network where the hubs have a smaller degree, thus favouring contagion. Finally, the critical threshold is slightly bigger for minority sources, suggesting again a bias in the information propagation. We conclude that the order of the transition is maintained, but the location of the transition point is strongly affected by $h$.  \\
We have also analyzed the dependency of the critical threshold on the number of nodes. Panel \ref{fig2}b shows that the size of the network changes the height of the curve, but not its shape. This implies that the dependency on the homophily parameter is robust with respect to size. Furthermore, in panel \ref{fig2}c we also observe that, regardless of the homophily, $T_c$ tends to zero as $N \to \infty$. This is a general phenomenon in scale-free networks \cite{watts_simple_2002}, because the higher degree of the hubs inhibits the contagion process.

\subsection*{Hybrid Contagion}
Even though for Complex Contagion we found that $h$ strongly affects  $IT$, the dependency on the source and target groups was minimal and lacked the strong biases -like echo chambers \cite{cinelli_echo_2021}- found in real-world networks. This discrepancy between model and data can be attributed to the "all-or-nothing" behavior of the system. Motivated by this, we propose a \textit{Hybrid Contagion} model (HC). In this new version, Simple Contagion represents viral transmission of information among nodes with sympathy towards the source of the information -for example, spreading of information is easier between individuals belonging to the same group-, whereas Complex Contagion models skeptical individuals that require multiple exposures to become convinced - e.g. when information comes from an unreliable source. This creates a scenario where one group has more difficulties when trying to convince members of the opposite group, in alignment with situations in real-world social networks.\cite{state2015}

\subsubsection*{Description of the model}
Hybrid Contagion is the combination of Simple and Complex spreading into a unified model. In this way, nodes on the network can follow Simple or Complex Contagion depending on their group and on the source of information. Nodes following Complex Contagion (called \textit{complex nodes}) become informed when they have a fraction of informed neighbours bigger that the threshold, $T$. Nodes following Simple Contagion (\textit{simple nodes}) can become informed with just one informed neighbour, with probability $\lambda$. \\
To simplify the model, we will make the following three assumptions. Firstly, we assign to each group a contagion type.; i.e, all majority nodes are simple and all minority nodes are complex, or vice versa. Secondly, we assume that the group that initiates the contagion process is always the group that follows Simple Contagion. This has the justification that ideas spread more easily among individuals that belong to the same group. For this same reason, we finally assume that the simple nodes always become informed whenever they have at least an informed neighbour; this is, we set the infectivity equal to one. $\lambda=1$. With these simplifications, the only remaining control parameter of this model is the threshold $T$.

 \subsubsection*{Information transmission with Hybrid Contagion}
  \begin{figure}[ht]
\centering
\includegraphics[scale=0.8]{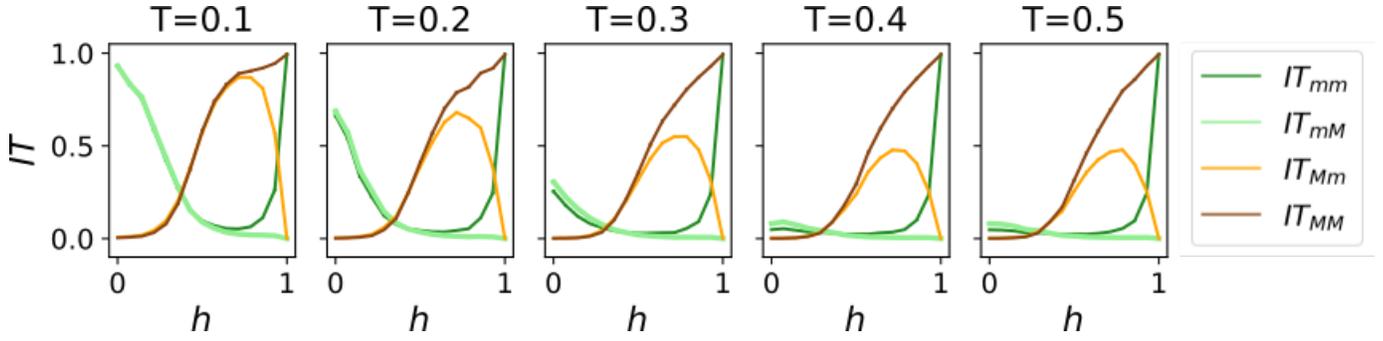}
\caption{ Information transmission observables as a function of the homophily parameter $h$, for Hybrid Contagion dynamics. Each column shows a different threshold $T$. The first subindex of each element of the legend indicates the source group, while the second subindex indicates the target group. Parameters used: $N=1000$ nodes, minority fraction $f_a=0.2$, $M=1000$ realizations.    }
\label{fig4}
\end{figure}

To investigate the predictions of this new model, we have calculated as before the information transmission for all combinations of source and target (figure \ref{fig4}).
This time, we observe not only a strong dependency on $h$, but also on the source and target groups (the latter only in the homophilic regime). In fact, the two possible source groups show opposite tendencies: IT from a minority source is enhanced in heterophilic networks, while IT from a majority source is enhanced in homophilic networks. These two opposite effects can be simultaneously explained in terms of one single observation: IT in Hybrid Contagion is favoured when the network hubs follow Simple Contagion. The effects of the threshold $T$ are also different depending on the source: increasing $T$ causes a strong drop in the IT with minority source, but the change is much weaker with a majority source. It is remarkable that, although $T$ is a parameter that only controls the behavior of complex nodes, the information transmission to simple nodes is affected too; in fact, in the heterophilic regime it is irrelevant whether the target node is simple or complex. This can be attributed to the high bipartivity of the network (most routes connecting two simple nodes cross a complex node). \\
Additionally, the source and target dependencies imply strong biases in the information transmission. In particular, IT from minority to majority becomes negligible for $h>0.5$: we have a lack of IT even when the network is connected. On the other hand, information transmission from majority to minority is nonzero. This asymmetry suggests the existence of some phenomenon similar to an echo chamber, which will be the main topic of the following sections. \\

\subsubsection*{Critical threshold for Hybrid Contagion}
\begin{figure}[ht]
\centering
\includegraphics[width=\linewidth]{HC_example_together.pdf}
\caption{Examples of Hybrid Contagion on BAh networks. Each network shows the final state of one simulation, where green nodes are susceptible and yellow ones are informed. The orange node is the source of the information. Simple nodes are denoted by circles and complex nodes by squares. a) When the source node belongs to the minority, Hybrid Contagion is limited. b) However, when the source node belongs to the majority, a wide range of cascade sizes is obtained for the same parameters. Parameters used: $N = 50$, $h = 0.7$, $f_a = 0.2$, $T = 0.2$.}
\label{aux3}
\end{figure}

As happened for Complex Contagion, a system following Hybrid Contagion shows a rich dynamics that is hard to understand by looking only at the IT. A simplified version of these rich dynamics is shown in fig. \ref{aux3}, for a threshold $T=0.2$. One sees that a minority source produces a small cascade (panel a), consisting mainly of nodes of the same group. On the other hand, a majority source produces cascades with a wide size range, from $70\%$ (panel b1) to $100\%$ (panel b2) of the nodes. In some cases, clusters of simple nodes get shielded from contagion due to the presence of a complex node. \\
\begin{figure}[ht]
\centering
\includegraphics[width=\linewidth]{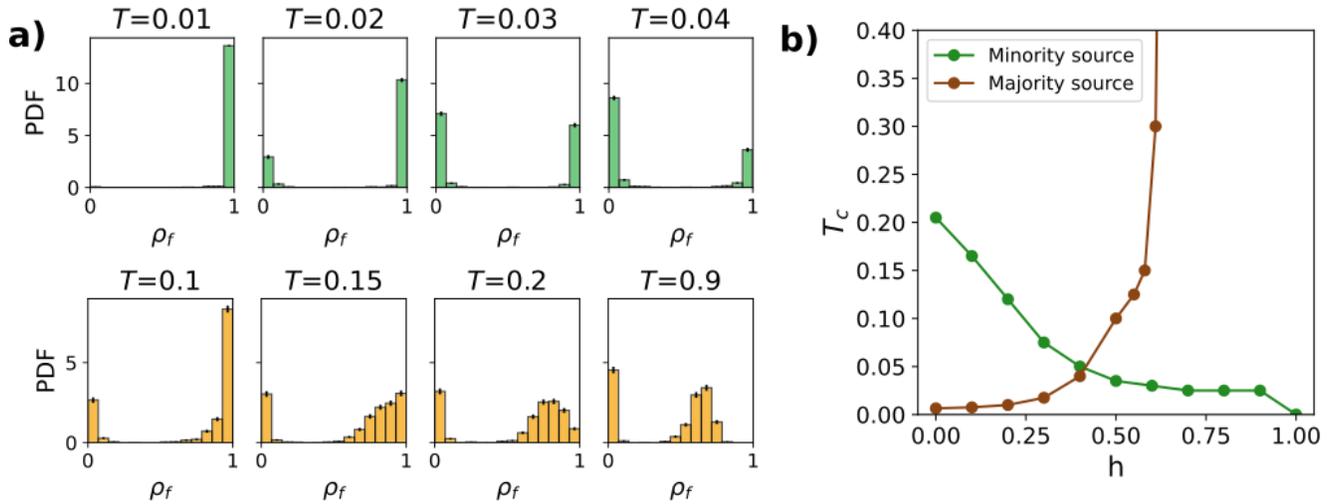}
\caption{Critical threshold for Hybrid Contagion dynamics. a): Histograms with the probability distribution of the final density of informed nodes, for a minority source (green histograms) and a majority source (orange histograms). The pdf is normalized so that the total area of each histogram equals one. The homophily parameter is $h=0.6$ b): Critical threshold as a function of the homophily parameter, for minority and majority sources. Other parameters used: $N=1000$ nodes, $M=1000$ realizations, $f_a=0.2$.    }
\label{fig5}
\end{figure}
To investigate this behavior in more depth and to study the nature of the transition, we calculate the pdf of the density of informed nodes for different values of the threshold $T$ (figure \ref{fig5}, panel a). As opposed to Complex Contagion, where differences between majority and minority source were minimal, now we see two different behaviors depending on the source group. In particular, for a minority source (upper panels of figure \ref{fig5}a), we obtain the "all-or-nothing" behavior typical of Complex Contagion. This totally changes when the source belongs to the majority , with a different form of the pdf that reflects the different behaviors of figure \ref{aux3}b. Moreover, the pdf does not change in shape for thresholds above $T=0.2$ (compare, for example, the histograms corresponding to $T=0.2$ and $T=0.9$ in the lower panels of figure \ref{fig5}a). Nevertheless, both histograms show a discontinuous jump of the global maximum. All these phenomena could be indicators of a hybrid transition \cite{baxter_heterogeneous_2011, lee_hybrid_2016}, where one of the maxima of the pdf varies continuously but the global maximum changes discontinuously at a certain threshold value $T_c$.
 \\
When we plot the change of the critical threshold $T_c$ as a function of $h$ (panel \ref{fig5}b), we observe opposite tendencies depending on the source groups: for minority sources, $T_c$ decreases with $h$, while for majority sources, it increases. More interestingly, for a majority source, $T_c$ diverges around $h=0.63$. Above this value, only the phase corresponding to information transmission exists. In short, for a majority source and sufficiently homophilic networks, the information is always able to propagate through the network, regardless of the threshold.

\subsection*{Echo chambers and other IT biases} 
 \subsubsection*{Measuring echo chambers and IT biases}
 As discussed in the previous section, the Hybrid Contagion model shows a rich behavior that cannot be easily understood by simply measuring the four IT observables. In particular, Hybrid Contagion exhibits \textit{information transmission biases} (IT biases), which can be defined as dependencies of information transmission on the source and target groups. To be able to classify these biases, as well as to determine their strength, we propose the following set of \textit{bias variables}:
\begin{subequations}  \label{social_parameters}
 \begin{align}
        \overline{IT}&=\frac{1}{4}(+IT_{mm}+IT_{mM}+IT_{Mm}+IT_{MM}) \\
    B_{E}&=\frac{1}{2}(-IT_{mm}-IT_{mM}+IT_{Mm}+IT_{MM}) \\
    B_{R}&=\frac{1}{2}(-IT_{mm}+IT_{mM}-IT_{Mm}+IT_{MM}) \\
    B_{EC}&=\frac{1}{2}(+IT_{mm}-IT_{mM}-IT_{Mm}+IT_{MM}) 
\end{align}
\end{subequations}
Notice that, since $0<IT_{ab}<1$, we have $0<\overline{IT}<1$ and $-1<B_{E},B_{R},B_{EC}<1$.  \\
The previously used IT observables focused on specific groups, and as a consequence they lacked information about the information propagation on the network as a whole. In contrast, the bias variables are able to distinguish global aspects of the transmission process, e.g. the receptivity of the nodes towards certain types of information. The first one, $\overline{IT}$, is the \textit{mean information transmission} over the four possible combinations of source and target. It is high whenever the network transmits information effectively regardless of who emits and who receives the information. Secondly, $B_{E}$ is the \textit{emissivity bias}. If it is bigger than zero, it indicates that the information that starts in the majority propagates more effectively than starting in the minority, regardless of the target. On the contrary, $B_{E}<0$ indicates that information starting in the \textit{minority} propagates more effectively. Thirdly, $B_{R}$ is the \textit{receptivity bias}. $B_{R}>0(<0)$ indicates that the majority has a bigger (smaller) probability of receiving information than the minority. Finally, we define $B_{EC}$ as the \textit{echo chamber bias}. It estimates whether information propagates only between nodes of the same group or not. If it is close to one, information only propagates inside of each group; while if it is negative, it indicates the existence of an "anti echo chamber": information only flows between nodes of the opposite groups. \\
Armed with these new variables, we can easily determine the biases of an information transmission process. One just needs to measure the IT parameters $IT_{ab}$ from a given data set and substitute them into eqs. \eqref{social_parameters}. Eqs. \eqref{social_parameters} also gives an objective definition of echo chamber: a social network shows an echo chamber if the bias $B_{EC}$ of an information transmission process is larger than zero. Finally, one should also note that \textit{any} bias in the information transmission between two groups can be expressed as a linear combination of the three biases (plus a possible contribution from $\overline{IT}$). In other words, any bias can be decomposed into a mixture of $B_E, B_R$ and $B_{EC}$.

\begin{figure}[ht]
\centering
\includegraphics[scale=0.8]{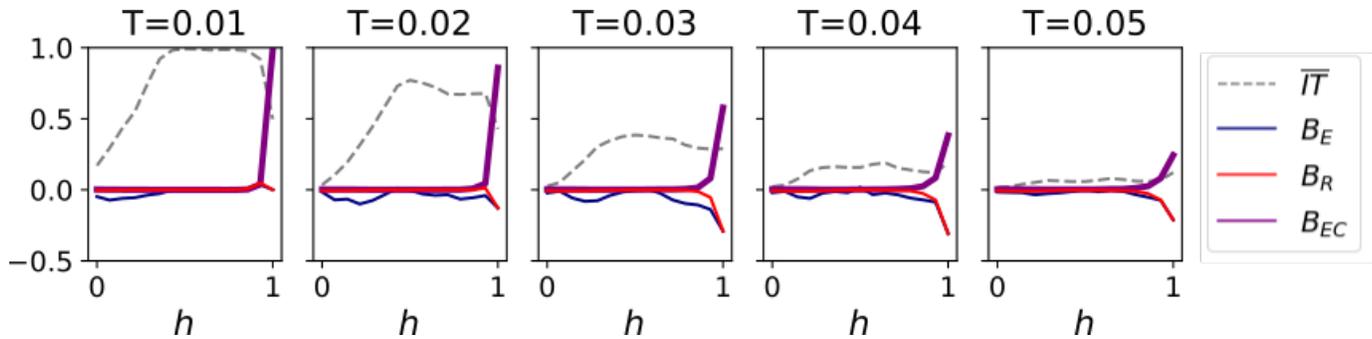}
\caption{Bias variables $\overline{IT},B_{E}$, $B_{R}$ and $B_{EC}$ as a function of the homophily parameter $h$, for Complex Contagion dynamics. The simulation parameters are the same as in figure \ref{fig1}b.
     }
\label{fig3}
\end{figure}
As an example of the usefulness of this new framework, we will analyze the results of Simple and Complex Contagion -already analyzed in fig. \ref{fig1}- in terms of these variables. As expected, Simple Contagion presents negligible biases $B_E, B_R, R_{EC}$ (not shown). A similar picture arises for Complex Contagion (fig. \ref{fig3}): only $\overline{IT}$ shows an important dependency on $h$; however, its behavior is the same that we saw in fig. \ref{fig1}. Additionally, the other three bias variables show an almost neutral behavior. Only significantly high values of $h$ lead to an increase in $B_{EC}$ and $|B_R|$ until $h=1$, where they reach their maximum due to the fragmentation of the network. These last results draw a clear picture: the transmission of information under Complex Contagion is sensitive to the homophily parameter and may be greatly inhibited, but it does not show strong biases.

 \subsubsection*{Echo chambers and other biases in Hybrid Contagion}
  \begin{figure}[ht]
\centering
\includegraphics[scale=0.8]{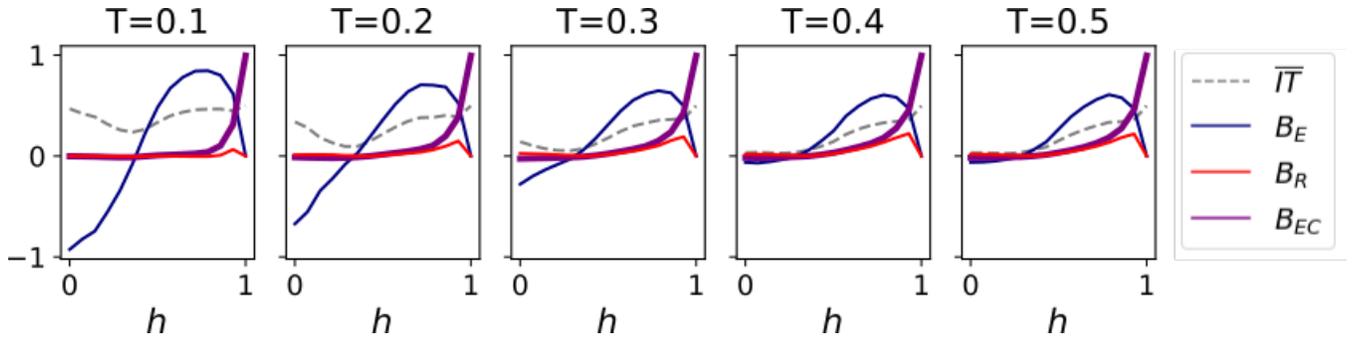}
\caption{Bias variables $\overline{IT},B_{E}$,  $B_{R}$ and $B_{EC}$ as a function of the homophily parameter $h$, for Hybrid Contagion dynamics. The simulations and parameters are the same as in figure \ref{fig4}. }
\label{fig6}
\end{figure}
 Unlike Complex Contagion, Hybrid Contagion shows important dependencies on both the source and target groups. The most relevant was the impossibility of transmitting information  from minority to majority for $h>0.5$. To quantify these dependencies, we plot the four bias variables for Hybrid Contagion in figure \ref{fig6}. \\
 Importantly, $\overline{IT}$ is smaller than one and practically constant with respect to $T$ and $h$, although it is minimum in the heterophilic regime. Additionally, the system has a strong emissivity bias, negative in the heterophilic regime and positive in the homophilic. Increasing the threshold reduces the absolute value of the emissivity in the heterophilic regime, with a limited effect in the homophilic one. The receptivity bias becomes also appreciable for $h>0.9$ and large thresholds. However, the most relevant observation is that the echo chamber variable is different from zero throughout \textit{all} the homophilic regime ($h>0.5$), especially for higher thresholds and for high homophily parameters.  \\
 Summing up, the combination of Simple and Complex Contagion (Hybrid Contagion) leads not only to strong emissivity biases for a wide range of $h$, but also to the emergence of echo chambers in the homophilic regime. Nevertheless, the average information transmission remains almost constant and is minimum in the heterophilic regime. This highlights the independence between the quantity and the lack of biases of transferred information: neither the presence of echo chambers and other biases implies that information transmission is hindered, nor a strong transmission of information guarantees that the information is bias-free.

\subsubsection*{Echo chambers in real-world networks}
   \begin{figure}[ht]
\centering
\includegraphics[scale=0.75]{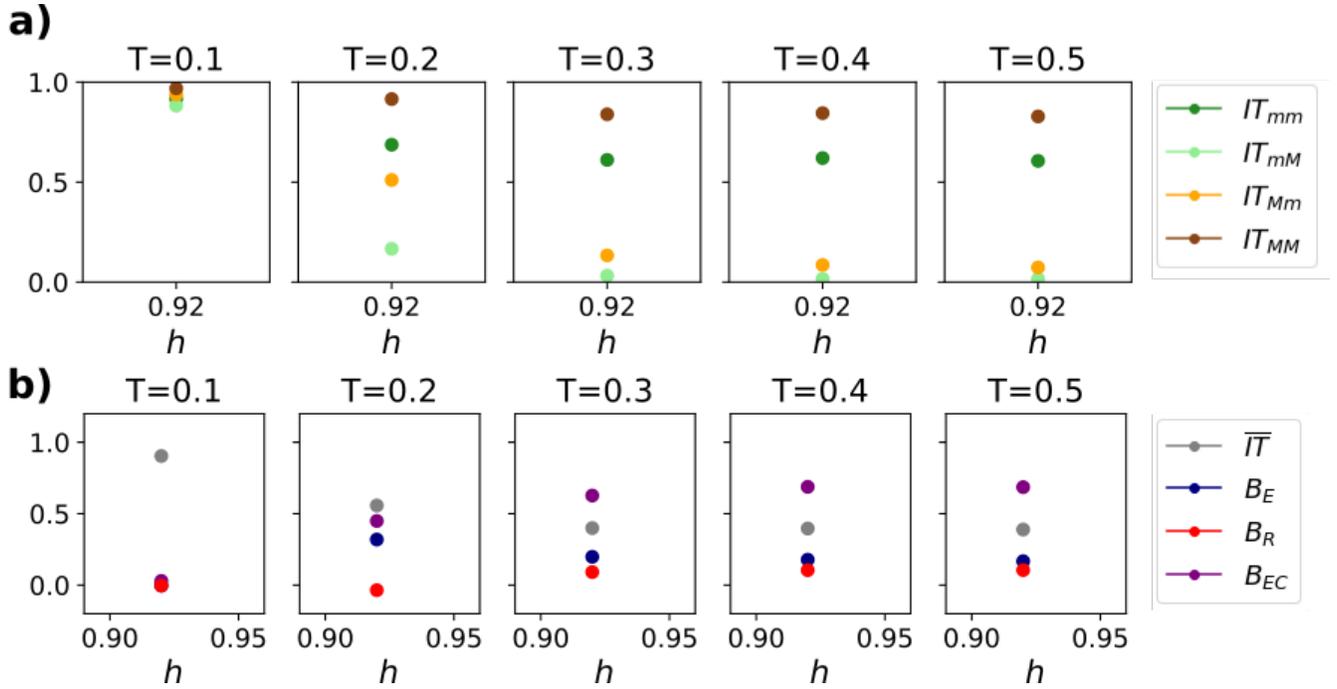}
\caption{a) Information transmission observables for Hybrid Contagion dynamics in a network of scientific citations between papers of the APS, for different thresholds $T$. The legend follows the same conventions as figure \ref{fig4}. Parameters of the network: $h=0.92$, $f_a=0.3177$. b) Bias variables corresponding to the IT observables of panel a.} 
\label{fig_APS}
\end{figure}
To put our results in a more solid ground, we test our results on a real collaboration network. We simulate Hybrid Contagion in a network of scientific citations between papers published in the journals of the American Physical Society (fig. \ref{fig_APS}) . This network was already used to study perception biases in \cite{lee_homophily_2019}. In particular, we focus on papers devoted to statistical mechanics, where majority and minority groups have been identified with Quantum Statistical Mechanics and Classical Statistical Mechanics, respectively. The network is highly homophilic, with an homophily parameter $h=0.92$. \\
Repeating our analysis regarding the IT observables and the bias variables, we obtain results that are fully consistent with the prediction in synthetic networks (figure \ref{fig4}). Indeed, the four IT observables show the same behavior as the synthetic curves for high homophily parameters ($IT_{MM}>IT_{mm}>IT_{Mm}>IT_{mM}$ in panel \ref{fig_APS}a. Surprisingly, we find high values of the echo chamber bias ($B_{EC}\approx 0.5$), even higher than in synthetic networks with the same parameter $h$ (panel \ref{fig_APS}b). A possible explanation is that  real-world networks have a higher clustering than our synthetic networks. This could favour intragroupal contagion and enhance the echo chamber bias.  \\
The remaining bias variables are also in agreement with our predictions in synthetic networks: the emissivity is positive, and information transmission remains high even for large thresholds. In summary, our results contribute to the understanding of the emergence of echo chamber biases in information transmission between different groups.

\section*{Discussion}
In this work, we have explored how information transmission (IT) in homophilic/heterophilic scale-free networks can be modelled, focusing on alterations of information transmission such as the emergence of echo chambers. To achieve this, we have analyzed three different models and proposed a decomposition of information transmission that allowed a straightforward quantification of the presence of biases. Starting from a structural model able to generate networks with tunable level of homophily \cite{karimi_homophily_2018}, we analyzed Simple Contagion by employing a slightly modified version of the SIR dynamics; Complex Contagion, and finally a Hybrid model in which the spreading between two nodes changes depending on whether they belong to the same or to different groups. \\ 
Our main conclusion is that the choice of the dynamical model greatly influences both the average information transmission and the emerging biases. In particular, we find that Simple Contagion leads to negligible biases and a minimal dependency on the homophily parameter $h$, whereas Complex Contagion shows strong dependencies on $h$, with high average information transmission in the homophilic regime. These differences are also reflected in the Complex Contagion transition between the informed and uniformed regimes. In particular, the threshold for the transition is higher in the homophilic regime, leading to an enhancement of the information transmission. The strong variability of the mean IT when changing $h$ is not correlated with strong biases in information transmission. In fact, most biases only appear as the network becomes disconnected, with the exception of a slight emissivity bias for a wider range of $h$. \\
A richer phenomenology is found in the Hybrid Contagion model, in which information transmission has an important dependency on $h$ and follows opposite trends when changing the source of the information. In particular, information originated in a minority node spreads easily in the heterophilic regime, while in the homophilic regime the transmission is dominated by information originated in the majority. 
Thus, interchanging the source group affects not only the behavior of the transition -with a divergence of the majority source critical threshold at $h\approx0.63$-, but also the qualitative behavior of the pdf, with hints of a possible hybrid transition.
As opposed to Complex Contagion, the dependencies on the source and target groups cause the appearance of stronger IT biases, not only in the emissivity and receptivity of information, but also in the emergence of echo chambers, where information starting in the minority fails to reach the majority groups. The echo chamber bias is different from zero for any $h>0.5$, which implies that intragroup communication is favoured and intergroup communication is hindered in all the homophilic regime. Moreover, even though for $h<1$ the network is still connected, the echo chamber bias reaches a significant value ($B_{EC}\approx0.5$).
When analyzing a citation network between papers of the APS, we find even stronger biases, highlighting the relevance of our results for real-world scenarios. \\
On a more general note, our study points out three important factors when analyzing information transmission. Firstly, that homophily and heterophily play a key role in how well information is transmitted and which biases appear. Secondly, that the quantity of transmitted information is not necessarily correlated with lack of biases: our analysis showed that models with low average information transition can be free from biases (like Complex Contagion for high threshold parameters), whereas models with high mean information transmission can show strong biases (Hybrid Contagion). Thirdly, biases in information transmission are not limited to echo chambers. Other biases (such as different levels of emissivity) can play a comparable role and affect the transmission of knowledge in our society \cite{fricker_epistemic_2007, barbera_tweeting_2015}. In summary, we believe that our decomposition of information transmission into an average value plus three distinct biases can help clarify complicated information transmission patterns in real data. \\
All the presented models show an important limitation: our group category is binary. In general, splitting of society's complexity into just two groups is too reductive, one clear example being the aggregation of political viewpoints into "left" and "right" groups. In this sense, a generalization of the BAh model to a continuum of groups would be helpful to better understand how individuals can gradually transition between groups, and may cause unexpected behaviors in the bias magnitudes. Another variation of these models could also incorporate a coevolving network in which nodes rewire to maximize the number of neighbors with their same opinion. This rewiring is known to lead to network polarization, echo chambers and ultimately fragmentation \cite{vazquez2008,saeedian2020,tokita_polarized_2021}, but its influence on the other biases is unknown. Finally, the generalization of the model to multidimensional topic spaces could help in the understanding of how ideologies form \cite{baumann_emergence_2021}.  \\
In conclusion, we have shown that contagion models beyond Simple Contagion can exhibit information transmission biases, including echo chambers. We hope that this works provides awareness about information transmission biases and the simple mathematical tools able to quantify them, so that further research can better understand how they emerge and ultimately overcome them.

\section*{Methods}
\subsection*{Measuring information transmission}
As mentioned in the main text, the final density of informed nodes within a group $g_t$, when taking the seed node within a group $g_s$, coincides \textit{on average} with the probability of information transmission from a source from $g_s$ to a target from $g_t$. Here we present a proof of this equivalence. \\
Let $N_g$ be the number of nodes belonging to group $g$, and let $I_g$ the number of informed nodes of group $g$ in the final (absorbing) state. We define the final density of informed nodes within group $g$ as $\rho_{fg}=\frac{I_g}{N_g}$. On the other hand, the \textit{average} probability that information spreads from a given source to a given target, $P_{tr}$, when choosing source and target nodes randomly within the source and target groups, is:
\begin{align}
    P_{tr}=\textrm{Pr}(\textrm{target is informed }|\textrm{ contagion starts at source})=\frac{\textrm{No. targets that are informed }}{\textrm{No. of possible targets}}=\frac{I_g}{N_g}=\rho_{fg}
\end{align}
Thus, we have shown that measuring the information transmission observables can be reduced to measuring the final density of informed nodes. \\
In the following, we describe the exact simulation procedure to measure $\rho_{fg}$: \\
1. Set the source and target groups. \\
2. Generate a network with the Barabasi-Albert-homophily model.\\
3. Select a source node and a target node randomly, with the constraint that they belong to the source and target groups respectively. \\
4. Mark the source node as informed. Simulate a contagion process with the corresponding model (Simple, Complex or Hybrid Contagion). \\
5. Once the absorbing state is reached, measure the number of nodes belonging to group $g$, $N_g$,  and the number of informed nodes of group $g$, $I_g$, to get the final density of informed nodes of group $g$, $\rho_{fg}$. \\
6. Repeat steps 2-5 for several networks and source and target nodes, and find the average final density of informed nodes $\rho_{fg}$.

\subsection*{Empirical network}
To test the validity of our results in real networks, we performed a simulation in a network with high homophily: a network of scientific citations of the APS \cite{karimi_homophily_2018, lee_homophily_2019}. The nodes in the network correspond to individual scientific papers related to statistical mechanics, and each link corresponds to a citation between two of them. We disregard the directional nature of the links, since the BAh model is designed for undirected networks. To ensure that contagion is possible, we only take the largest component and disregard all the small components. We select the minority and majority groups based on identifiers from the Physics and Astronomy Classification Scheme (PACS). In particular, the minority group corresponds to papers devoted to Classical Statistical Mechanics (CSM) and the majority group to Quantum Statistical Mechanics (QSM). \\
Taking these considerations into account, we obtain a network with 1281 nodes and 3064 links. From these nodes, 407 belonged to the minority group and 874 to the majority. The minority fraction is thus $f_a=0.32$, and the homophily parameter is $h=0.92$. The homophily parameter was estimated using the procedure described in \cite{karimi_homophily_2018}. 




\newpage
\bibliography{DSMM_fv}

\section*{Acknowledgements}
FDD, MSM and SM acknowledge support from the Spanish Agency of Research (AEI) through Maria de Maeztu Program for units of Excellence in R\&D (Grant MDM-2017-0711 funded by MCIN/AEI/10.13039/501100011033). In particular, FDD thanks financial support
MDM-2017-0711-20-2 and FSE invierte en tu futuro. SM acknowledges funding from the project PACSS RTI2018-093732-B-C22 of the MCIN/AEI /10.13039/501100011033/ and by EU through FEDER funds (A way to make Europe).

\section*{Author contributions statement}

FDD, MSM and SM conceived the models. FDD performed the numerical simulations and analyzed the data. MSM. and SM supervised the work. All authors contributed in the writing and revision of the manuscript. 

\section*{Additional Information}

\textbf{Competing interests:} The authors declare no competing  interests.


\end{document}